\documentclass[draft]{agujournal2019}
\usepackage{url} 
\usepackage{lineno}
\usepackage[inline]{trackchanges} 
\usepackage{soul}

\usepackage{graphicx}
\graphicspath{{./}}
\usepackage{setspace}
\usepackage{lineno}
\usepackage{soul}
\usepackage{amsmath}

\usepackage{tikz}
\usepackage{color, colortbl}
\definecolor{Gray}{gray}{0.9}
\graphicspath{{./figs/}{}}
\usepackage{siunitx} 
\draftfalse

\begin{document}

%
%


\title{Thermal disequilibrium during channelized melt-transport: Implications for the evolution of the lithosphere-asthenosphere boundary}

%
%




\authors{Mousumi Roy}


\affiliation{1}{Department of Physics and Astronomy, University of New Mexico}




\correspondingauthor{Mousumi Roy}{mroy@unm.edu}




\begin{keypoints}
\item A 1D-column model suggests thermal disequilibrium during melt rock interaction may modify the base of continental lithosphere over geologic time scales
\item Episodic melt-infiltration at the lithosphere-asthenosphere boundary may build a steady-state thermal re-working zone (TRZ) in the lowermost lithosphere
\item The spatial and temporal scales for establishment of the TRZ are comparable to those inferred for the degradation of continental lithosphere in intraplate settings
\end{keypoints}

%
%

%
%


\begin{abstract}
This study explores how thermal disequilibrium during melt-infiltration and melt-rock interaction may modify the continental lithosphere from beneath.  
Using an idealized 1D model of thermal disequilibrium between melt-rich channels and the surrounding melt-poor material, I estimate heat exchange across channel walls during channelized melt transport at the lithosphere-asthenosphere boundary (LAB).
%
%
For geologically-reasonable values of the volume fraction of channels ($\phi$), relative velocity across channel walls ($v$), channel spacing ($d$), and the timescale of episodic melt-infiltration ($\tau$), model results suggest disequilibrium heating may contribute $>$ $10^{-3}$ W/m$^3$ to the LAB heat budget.
During episodic melt-infiltration, a thermal reworking zone (TRZ) associated with spatio-temporally varying disequilibrium heat exchange forms at the LAB.
The TRZ grows by the transient migration of a disequilibrium-heating front at material-dependent velocity, reaching a maximum steady-state width $\delta\sim$ $\left[\phi vd^{-2}\tau^{2} \right]$.
The spatio-temporal scales associated with establishment of the TRZ are comparable with those inferred for the migration of the LAB based on geologic observations within continental intra-plate settings, such as the western US.
%
\end{abstract}

%
%

%


%
%
%
%

\section{Introduction}
There is growing speculation, based on observations in a wide range of tectonic settings, that melt-infiltration may profoundly alter and occasionally destabilize continental lithosphere \cite{HopperMalawi20,wenkerbeaumont17,plankforsyth16,roy2016,wangplume15, menzies2007, carlson2004,  gao2002, reilly2001are}. 
The processes by which such alteration may occur, however, remain elusive.
This study explores thermal disequilibrium during melt-infiltration and melt-rock interaction as a means of shaping the continental lithosphere from beneath.  
Specifically, I explore thermal disequilibrium between melt-rich channels and surrounding material as a process to heat and modify the continental lithospheric mantle (CLM). 
A central idea explored here is that melt extraction pathways in the CLM, above the lithosphere-asthenosphere boundary (LAB) may involve a significant degree of disequilibrium heat exchange.

This study is inspired by evidence for the role of thermal disequilibrium from detailed field-based, petrologic, and geochemical  the Lherz and Ronda peridotite massifs \cite<e.g.,>[summarized in SI, Text S1]{soustelle2009, leroux2007lherz}.
I build on the idea that melt-rock interaction in the lower CLM may be characterized by zones with steep, transient thermal gradients as observed in these massifs \cite<e.g.,>[Text S1]{soustelle2009}.
Additionally, this work is motivated by observations from the western US, which has undergone extensive magma-infiltration in Cenozoic time.  
Pressures and temperatures of last equilibration of Cenozoic basalts consistently point to depths that are at or below the LAB \cite{plankforsyth16}, suggesting that melt transport from those depths upward through the lower CLM occurs in thermal disequilibrium.
In the Big Pine volcanic field, for example, the inferred depth of the LAB decreases by $>$10 km in a timespan of $<$1 Myr, suggesting that the processes associated with this migration may be transient.
More recently, Cenozoic melt- or fluid-enhanced thinning of the CLM in the western US has also been inferred from geochemical and isotopic data from volcanic rocks \cite{farmerTaTh20}.
Motivated by these observations, a primary goal of this work is to quantify the role of transisent, disequilbirium heating by infiltrating channelized melt as a mechanism for modifying the LAB and the lowermost CLM.

Typically, melt-transport in grain-scale percolative flow is assumed to occur in thermal equilibrium \cite<e.g.,>{mckenzie1984}.
Thermal disequilibrium during melt transport is expected to become important, however, as the degree of channelization and the relative melt-solid velocity increases  \cite<e.g.,>{schmelingdiseqm18}.
In this work, I am not concerned with the development of channel networks \cite<e.g. during reactive infiltration,>{aharonov95}, nor the processes that transport warmer-than-ambient melt from a sub-lithospheric melt-generation zone to the LAB.
Instead, the starting point of this study is the observation that high-porosity, melt-rich channels are an important part of melt-rock interaction in the lowermost lithosphere both in oceanic settings \cite<e.g.>{liuliang19} and the CLM \cite<e.g.>{soustelle2009, leroux2007lherz}.  
Therefore, I focus on the implications of significant thermal gradients between melt-rich channels and their surroundings \cite<e.g.,>{soustelle2009}.
%
Although others have also argued for the important role of thermal disequilibrium in melt-rock interaction \cite{kellersuckale19, wallnerschmeling16, schmelingdiseqm18}, this study provides a quantification of the role of thermal disequilibrium at the LAB based on observational constraints discussed above (also SI, Text S1). 

Using a simple 1D model, this study places constraints on the likely contribution of thermal disequilibrium to the heat budget at the LAB.
This work abstracts the complex geometry of the melt-rock interface and therefore differs from previous (more complete) descriptions of disequilibrium heat exchange  \cite{wallnerschmeling16, schmelingdiseqm18, kellersuckale19}.
Similar to reactive transport models that use a linear driving term for chemical disequilibrium \cite<e.g.,>{hauri97a, bo18traceelem}, the 1D models below assume a linear thermal driving term \cite<e.g.,>{schumann1929heat, kuznetsov94, spiga81}.
In the following section, the basic results of the 1D model are presented, followed by a discussion of their limitations and implications. 
Although idealized, the models are a fruitful way to assess the temporal and spatial scales over which thermal disequilibrium can play a role in warming and therefore weakening the lowermost CLM.
The first-order estimates of the rates and spatial scales of disequilibrium heating from the models are compared to geologic observations within the western US, specifically geochemical and petrologic evidence for the upward migration of the LAB during Cenozoic melt-rock interaction at the base of North America \cite{plankforsyth16, farmerTaTh20}. 

\section{Model of disequilibrium heat transport}
A simple, 1-D theory of heat exchange in packed porous beds is given by \citeA{schumann1929heat}, where fluid moves within the pores of a matrix of solid grains.
Here, the thermal evolution of the system is governed mainly by heat exchange across the solid-fluid interfacial surface.   
This heat exchange is assumed to dominate over thermal dispersion and axial conductive heat fluxes both within the moving fluid and in the surrounding regions.  
Additionally, heat exchange is assumed to be linearly proportional to the local temperature difference between solid and fluid.
These arguments lead to coupled equations for the temperature of the solid matrix, $T_s$, and within the fluid, $T_f$ \cite{schumann1929heat}:
\begin{linenomath*}
	\begin{equation}
		\frac{\partial{T_f}}{\partial{t}}+ v\frac{\partial{T_f}}{\partial{x}}=-\frac{k}{\phi c_f} (T_f - T_s)=-k_f(T_f - T_s)
		\label{eq:Tf}
	\end{equation}
	\begin{equation}
		\frac{\partial{T_s}}{\partial{t}}=\frac{k}{(1-\phi)c_s} (T_f - T_s)=k_s(T_f - T_s)
		\label{eq:Ts}
	\end{equation}
\end{linenomath*}
%
where $\phi$ is a fluid volume fraction, and $c_f$ and $c_s$ are the heat capacities per unit volume at constant pressure, so $c_f = c_{p\it {fluid}}\rho_f$ and $c_s = c_{p\it {solid}}\rho_s$.  
Note that the geometry of the solid-fluid interface is not treated in detail, but is idealized in the volume fraction, $\phi$ and in the fluid-solid heat transfer coefficient, $k$.  
This 1-D model has been investigated in numerous previous studies and analytic solutions for Eqns \ref{eq:Tf} and \ref{eq:Ts} have been derived for a number of limiting cases, particularly for large $k$ \cite{spiga81, kuznetsov94, kuznetsov95,kuznetsov95b,kuznetsov96}.

Here I present a re-interpretation of the equations above and of the heat transfer coefficient, made possible because the geometry of the interfacial surface is not explicitly specified.
Instead of considering fluid moving in pores between solid grains, the system of equations above may be used to describe thermal disequilibrium between material within high-porosity channels and outside channels.
In other words, here ``fluid" is interpreted to be in-channel material and ``solid" is material outside channels (for simplicity, I retain the subscripts $f$ and $s$ as above).
The velocity $v$ is therefore an average relative velocity across channel walls.
This ``coarse-graining" of the model must also be accompanied by an appropriate reinterpretation of the heat transfer coefficient, $k$, but solutions of Eqns \ref{eq:Tf} and \ref{eq:Ts} (particularly analytic solutions in limiting cases) above may be exploited. 
%

%
%
%

\begin{figure}[!htb]
\centering
\includegraphics[trim =30mm 50mm 30mm 50mm, clip, width=0.5\textwidth]{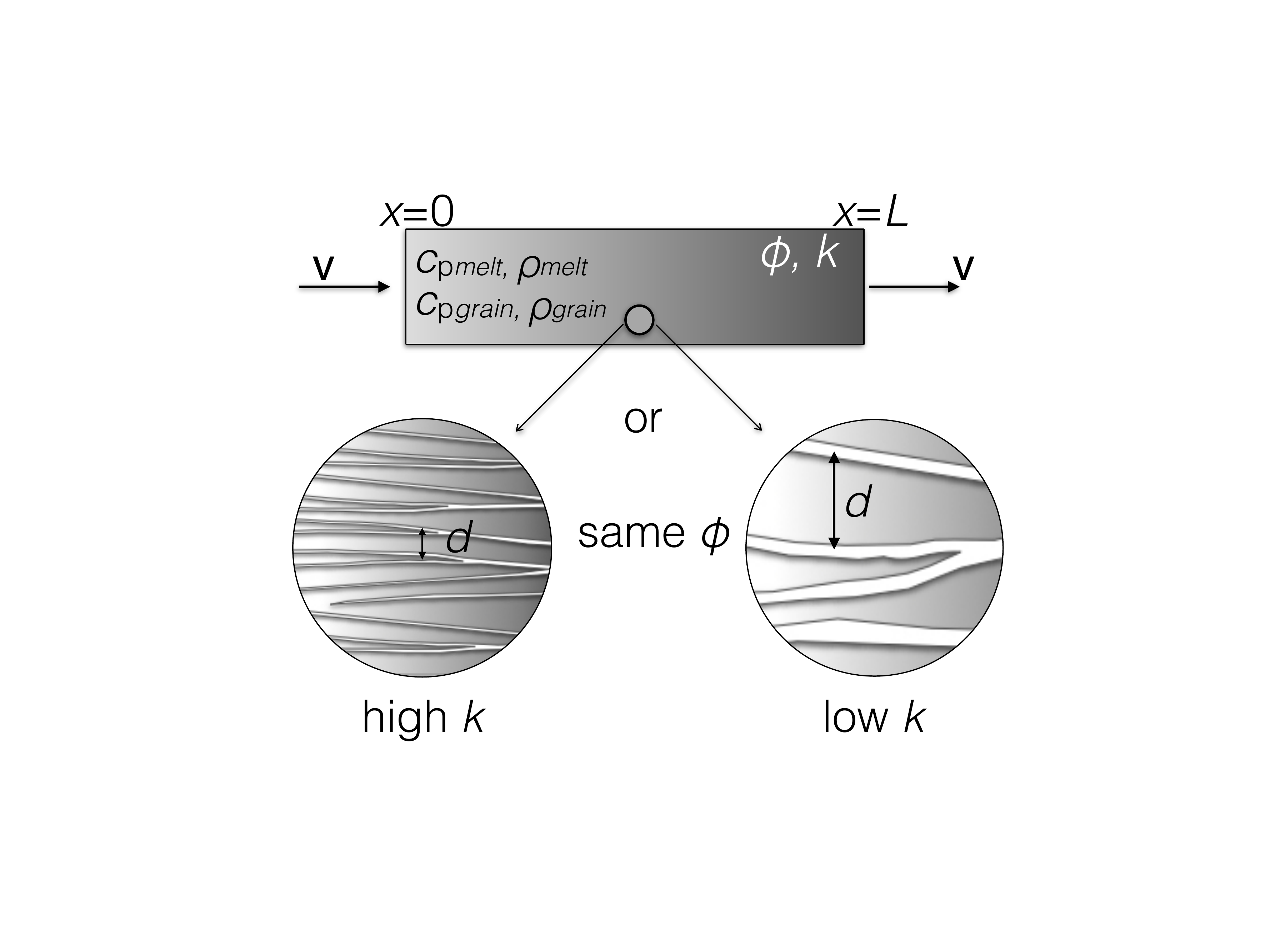}
\caption{Cartoon of 1D model with parameters such as: specific heat capacities ($c_p$) and densities ($\rho$), in-channel velocity $v$, channel volume fraction, $\phi$, and the heat transfer coefficient $k$ (SI, Table S1 for values). The heat transfer coefficient $k$ is a function of the geometry of the channels and scales as $d^{-2}$ , where $d$ is the channel spacing (see SI, Text S3); large $k$ corresponds to large channel wall area per unit volume (e.g., small $d$) and vice versa. }
\label{fig:model}
\end{figure}

The reinterpreted model is applied to a semi-infinite domain where fluid transport occurs in high-porosity channels aligned along one dimension (Figure \ref{fig:model}).  
The channels are assumed to occupy a constant volume fraction, $\phi$, within which material moves with a constant (average) velocity $v$ relative to the surrounding stationary material outside the channel (volume fraction $1-\phi$). 
%
%
The model domain may be thought of as co-moving with the reference frame of material outside the channels. 
Because of the assumptions built-in to the \citeA{schumann1929heat} approach, the results below are applicable to physical situations where: transport is in dominantly in the along-channel direction; heat exchange across channel walls dominates over conduction within channels and within walls; and any motion of material outside channels is steady.

The model assumes that the channel geometry is unchanging within the domain.  
The detailed geometry of the channel walls (the relevant interfacial surface here) is not specified but is parametrized by the heat transfer coefficient, $k$ (Figure \ref{fig:model}). 
Therefore, $k$ is a proxy for the geometry of the channel wall interface, namely the wall area per unit volume, controlled by the spatial scale of channelization, $d$ (SI, Text S3). 
As illustrated in Figure \ref{fig:model}, a large value of $k$ may represent efficient heat exchange as in the case of many channels separated by a small distance.  
Conversely, a low value of $k$ would represent inefficient exchange, as in the case of a larger characteristic length scale between the channels. 

The two independent factors on the right hand sides of Eqns \ref{eq:Tf} and \ref{eq:Ts} specify the timescales of heat exchange within channels, $1/k_f = \phi c_f/k$,  and outside channels, $1/k_s = (1-\phi)c_s/k$.
Instead of solid and fluid heat capacities per unit volume (heat capacitances) as in \citeA{schumann1929heat}, here $c_f$ and $c_s$ now represent an average heat capacitance within and outside channels, respectively.
If the material in-/outside of the channels is characterized by a grain-scale porosity, $\varphi_{in}$ or $\varphi_{out}$, then $c_f$ and $c_s$ may be written as the volume-average of the values for solid grains and melt (SI, Text S3) .
A characteristic length scale emerges out of the relative motion across channel walls, $v/k_f$.  
These characteristic length and timescales are used to non-dimensionalize Eqns \ref{eq:Tf} and \ref{eq:Ts} (SI, Text S2) and obtain the results presented below.
The behavior of the model is determined by five user-specified quantities: the heat transfer coefficient, $k$, channel volume fraction $\phi$, the heat capacitances, $c_f$, and $c_s$, and the relative velocity across channel walls, $v$ (Table S1; SI). 

\subsection{Heat transfer coefficient}\label{sec:k} Before discussing model results, I consider the meaning of the heat transfer coefficient, $k$, and the related constants, ${k}_s= k/(c_s(1-\phi))$ and ${k}_f = k/(c_f\phi)$ in Eqns \ref{eq:Tf} and \ref{eq:Ts}. 
Since $k_f$ and $k_s$ have dimensions of inverse time, $k$ represents the amount of heat transferred across channel walls per unit time, per unit volume, per unit difference in temperature \cite<in >[this exchange is across the solid-fluid interface]{schumann1929heat}.  

The factors that determine $k$ are explored in SI, Text S3, but in summary: for a given channel volume fraction, $\phi$, $k$ is strongly controlled by the length scale of channelization, parameterized by the channel spacing $d$.
Therefore, to decide on a range of $k$ values appropriate to the LAB, I turn to observations of the scale of channelization in exhumed portions of the lower CLM.
Structural, petrologic, and geochemical data from the Lherz Massif suggest that melt-rock interaction has driven refertilization of a harzburgite body into lherzolite \cite{leroux2007lherz, leroux08isotope}.  
In the field, the lherzolite bodies are separated from each other by distances of several tens of meters and this is also the spatial scale of isotopic disequilibrium between metasomatizing fluids and the harzburgite parent material \cite{leroux08isotope}.  
With this as a proxy for the spatial separation of fluid-rich channels, I choose a broad range for the relevant spatial scale of channelization, $d=10^{-1}$ to $10^2$ m (10 cm to 100 m channel spacings).
%
The corresponding range of the heat transfer coefficient in the models is therefore $k \approx 10^{-5}$ to $10^{3}$ W m$^{-3}$K$^{-1}$.  
In the following, material properties, channel volume fraction, in-channel velocity, and heat transfer coefficient are fixed for each calculation (Table S1; SI).  
\section{Results}
This section presents the primary findings of the 1D model in Eqns \ref{eq:Tf} and \ref{eq:Ts}. Two scenarios are considered: (1) the response to a step-function, discussed in detail in SI, Text S4; and (2) response to a sinusodal temperature perturbation. 

\subsection{Response to a step-function thermal perturbation} 
The domain is initially at steady-state in equilibrium at temperature $T_0$, $T_s=T_f=T_0$ and at $t=0$ the temperature of the fluid entering at the inflow, $x=0$, is perturbed so that, $T_f(x=0, t\geq0)=T_0+\Delta T$ (introducing $\Delta T$ as a temperature scale into the problem). 
This disturbs the initial steady state and starting at $t=0^+$ material at the inflow is no longer in thermal equilibrium with material in the domain. 
The primary finding is that temperature profiles within the domain exhibit a transition or disequilibrium zone, lagging behind the fluid front, but migrating inward into the domain  \cite{schumann1929heat}. 
\citeA{kuznetsov94} derives an analytic expression for the migration rate in the limit that the degree of disequilibrium is small, and the numerical models here extend this (SI, Text S4).

A key result is that the rate at which this zone migrates is independent of the heat transfer coefficient, $k$, but depends on material properties.
The location of maximum disequilibrium (maximum $T_{f}-T_{s}$, and therefore the greatest heat exchange) lags behind the fluid front and progresses inward into the domain at a rate given by Eqn \ref{eq:migrate}; SI, Text S4), based on models using a range of $\phi$ and $v$ values (SI, Table S1).
\begin{linenomath*}
\begin{equation}
 V_{diseqm} \approx  v\left (\frac{c_f}{c_s} \right ) \left ( \frac{c_f \phi}{c_f \phi + (1-\phi)c_s} \right )\label{eq:migrate}
\end{equation}
\end{linenomath*}
Although the rate of migration of the disequilibrium front is not a function the heat transfer coefficient, the characteristic width of this zone and the degree of disequilibrium within it are strong functions of $k$ (SI, Text S4). 

\subsection{Response to a sinusoidal thermal perturbation}
Fluid entering the domain is hotter than the ambient initial temperature, but now the thermal contrast varies sinusoidally, representing pulses of high temperature material in fluid- or melt-rich channels.
Sinusoidal thermal pulses introduce a new timescale into the problem: the period  $\tau$. 
The relevant timescale to compare $\tau$ to is $1/k_s$ is the longest response timescale in the domain, associated with the thermal response of the material outside channels. 
For the material parameters in Table S1 (SI), and  channel spacing of $d=10$ to 100 m, the characteristic response timescale $1/k_s \approx 1 $ to $100$ yr, which is short compared to the timescales of geologic events.

\begin{figure}[!htb] 
\centering
\includegraphics[trim =8mm 110mm 10mm 14mm, clip, width=\textwidth]{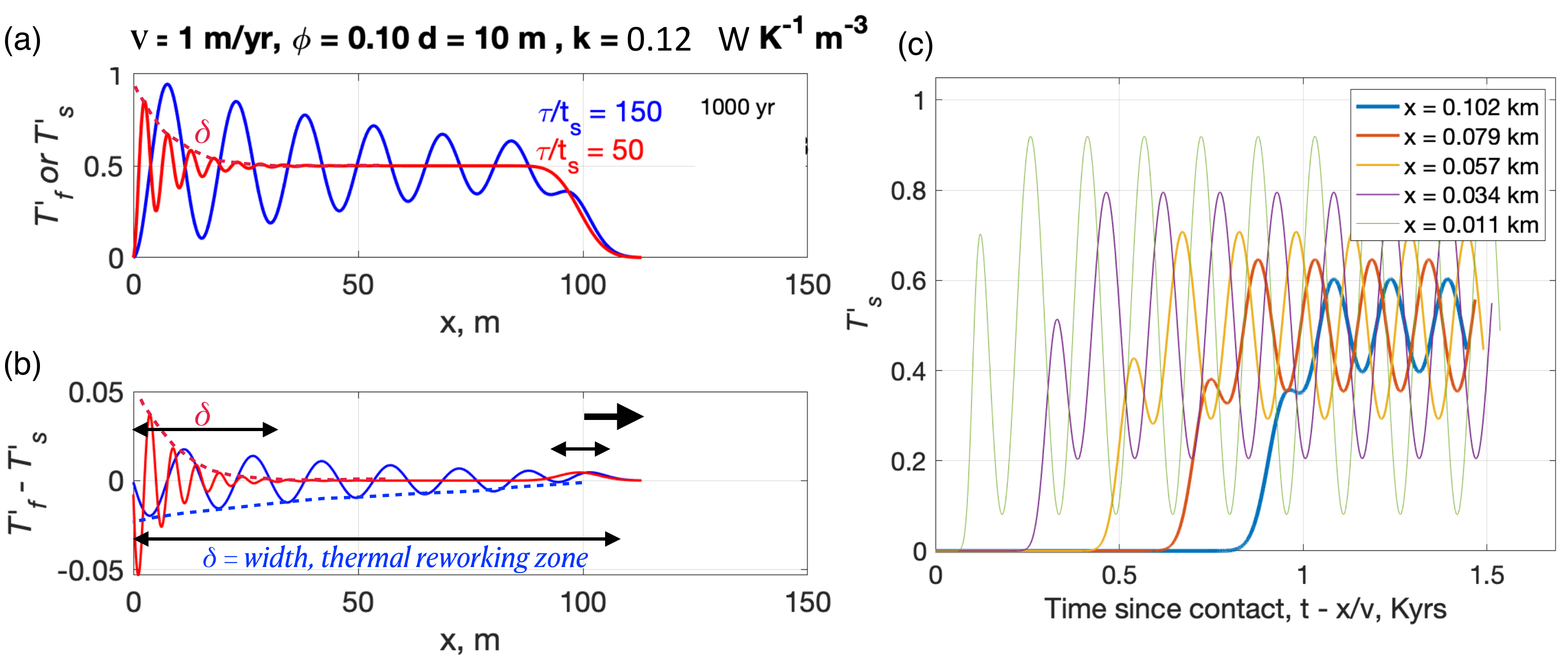}
\includegraphics[trim =8mm 7mm 10mm 8mm, clip, width=0.6\textwidth]{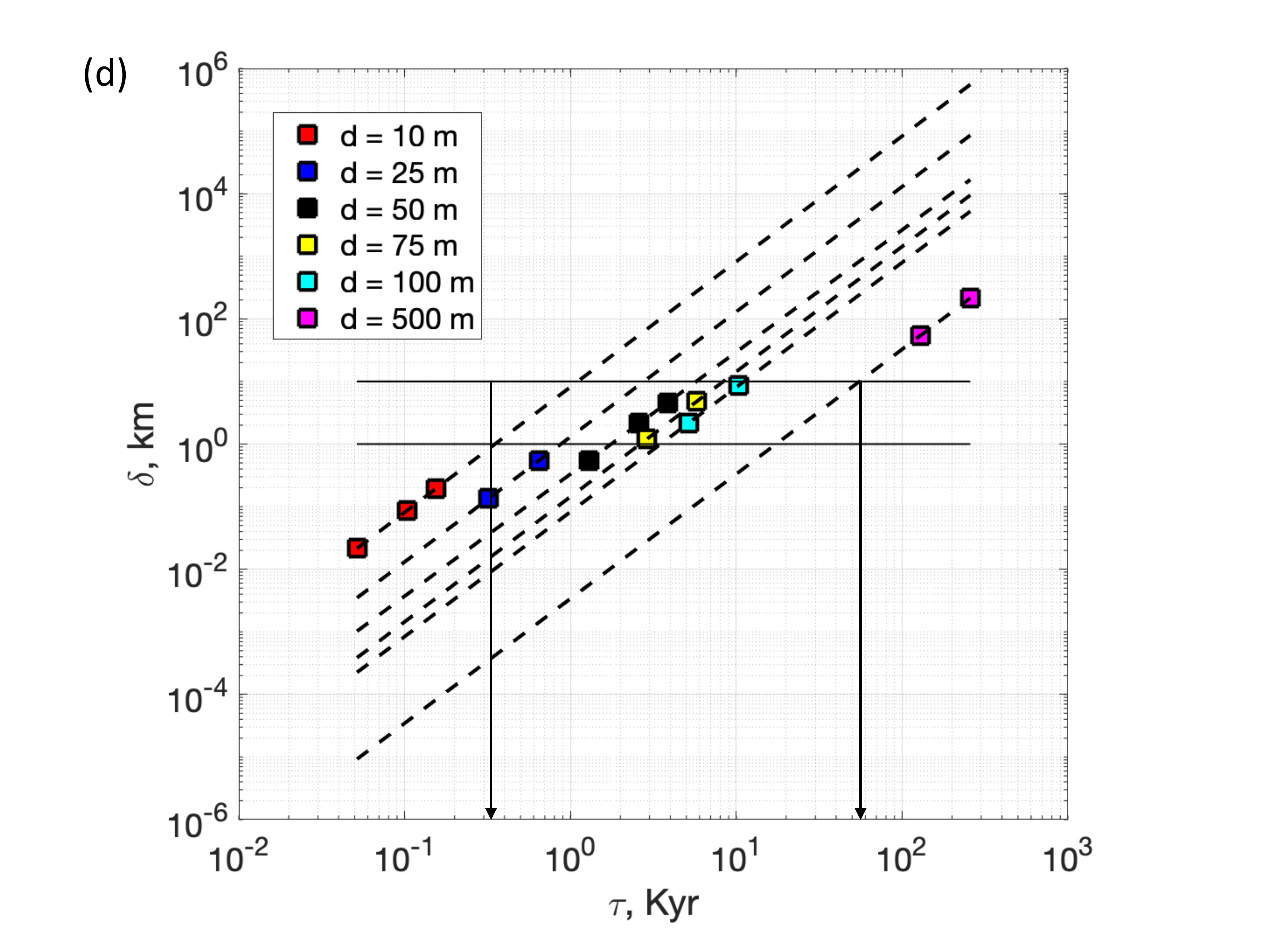}
\caption{(a) Normalized temperature profiles at time $t$=1 Kyr, in-channel $T'_{f}$ (solid lines), and out-of-channel $T'_{s}$ (dashes), for a calculation with in-channel velocity $v$=1 m/yr, channel volume fraction $\phi$=0.1, channel spacing $d$=10 m and heat transfer coefficient $k$ as indicated. For the chosen parameters, the response timescale is $1/k_s$=$t_s$=1 yr. Results are shown for two different thermal pulses in the incoming material with (normalized) oscillation periods: $\tau k_s$=50 (red), 150 (blue). The thermal reworking zone (TRZ) has spatial oscillations in $T'_{f}$ and $T'_{s}$, with amplitudes that decrease over a decay scale $\delta$, the width of the TRZ: $\delta$ is larger for longer period (blue) and shorter for shorter period (red).  (b) The degree of disequilibrium ($T'_{f}$$-$$T'_{s}$) is also oscillatory in the TRZ, with decaying amplitude over width $\delta$. (c) Temperature-time paths plotted at different distances from the inlet within $x$$<$$\delta$ for the case where $\tau/t_s$=150 in (a) and (b), illustrating temporal oscillations at each location within the TRZ. (d) Width of the TRZ, $\delta$, as a function of oscillation period $\tau$ and channel spacing $d$ as indicated. Dashed lines (slope 2) are the expected analytic scaling in Eqn \ref{eq:delta} and the squares indicate numerically derived values of $\delta$ obtained by fitting an exponential decay to the envelope of the ($T'_{f}$$-$$T'_{s}$) oscillations in the TRZ, e.g., in (b).  Thin horizontal lines are at $\delta$$=$$1$ and 10 km.  The vertical arrows indicate that for channel spacings of $d$$=$10 to 500 m, thermal pulses with periods of 200 yrs to 50 Kyrs will give rise to TRZ widths of 1 to 10 km depending on $d$.}
\label{fig:sinecase}
\end{figure}

Thermal pulses with periods that are long compared to $1/k_s$ penetrate farther into the domain than short period oscillations (Figure \ref{fig:sinecase}).
The results show that the non-dimensional period, $\tau k_s = \tau k/(1-\phi)c_s$, controls the length scale, $\delta$, over which thermal oscillations penetrate into the domain.
Therefore, periodic thermal perturbations that might represent melt infiltration pulses lasting $10^3$ to $10^6$ yrs will be characterized by a region of sinusoidally varying temperatures: a thermal re-working zone (TRZ) (blue curves in Figure \ref{fig:sinecase}a \& b).
The wavelength of these oscillations is set by the period $\tau$, $\lambda = v\tau$.
The penetration distance of the thermal oscillations, $\delta$, is the maximum width of the TRZ.
At short times, when $t \ll \delta/V_{diseqm}$, there is one zone of disequilibrium (the TRZ, bounded by the disequilibrium front).
At longer times, the TRZ widens to a maximum width, $\delta$, at time $\delta/V_{diseqm}$.
When $t \gg \delta/V_{diseqm}t$, there are two zones of disequilibrium: one stationary at the inlet (the TRZ), and the migrating zone discussed above that moves at $V_{diseqm}$ (Eqn \ref{eq:migrate}; red curves in Figure \ref{fig:sinecase}a \& b).

The oscillatory nature of temperatures inside the TRZ is illustrated in temperature vs. time paths within the domain at varying distances from the inlet (Figure \ref{fig:sinecase}c).  
The amplitude of the temperature oscillations decay with distance, but at each location in the TRZ the amplitude is constant, once oscillations are established (Figure \ref{fig:sinecase}c). 
As we might expect, the maximum width of the TRZ, $\delta$, is set both by the non-dimensional oscillation period, $\tau/t_s$ and by the heat transfer coefficient \cite<see also>{spiga81}, 
\begin{linenomath*}
\begin{equation}
 \delta = \left (\frac{c_f\phi v}{k} \right ) \left ( \frac{(\tau/t_s)^{2}}{4 \pi^{2}} \right )\label{eq:delta}
\end{equation}
\end{linenomath*}
noting that $k \sim d^{-2}$ (SI, Text S2), and $t_s = 1/k_s = c_s(1-\phi)/k$, the expression above suggests that, for fixed $v$, $\delta \sim (\tau/d)^2$ as confirmed by the numerical results (Figure \ref{fig:sinecase}d).



\section{Discussion}

The model above is highly idealized and therefore limited in its representation of the complexities of deformation and fluid-rock interactions within the Earth. 
In particular, the effective thermal properties and the geometry of the fluid-/melt-rich channels are abstracted into a single number, the heat transfer coefficient, $k$, strongly controlled by the channel spacing, $d$.
Sinuosity and other aspects of the geometry of channelization are abstracted and the details of processes at and below the scale of an average channel spacing, $d$, are ignored.
Instead, the focus here is on the effective behavior at mesoscopic spatial scales $\gg d$.
Even at these scales, we ignore spatial variations in transport, including variability in the channel volume fraction $\phi$, in-channel velocity $v$, and effective heat transfer coefficient $k$.
Time-dependent variability in transport, e.g., feedbacks due to possible phase changes during disequilibrium heating/cooling which would affect the geometry of the channels, are ignored \cite{kellersuckale19}.
Finally, this 1D model ignores the 3D nature of relative motion across channel walls even on the mesoscale ($\gg$$d$).

\begin{figure}[!htb] 
\centering
\includegraphics[trim =60mm 35mm 60mm 40mm, clip, width=0.65\textwidth]{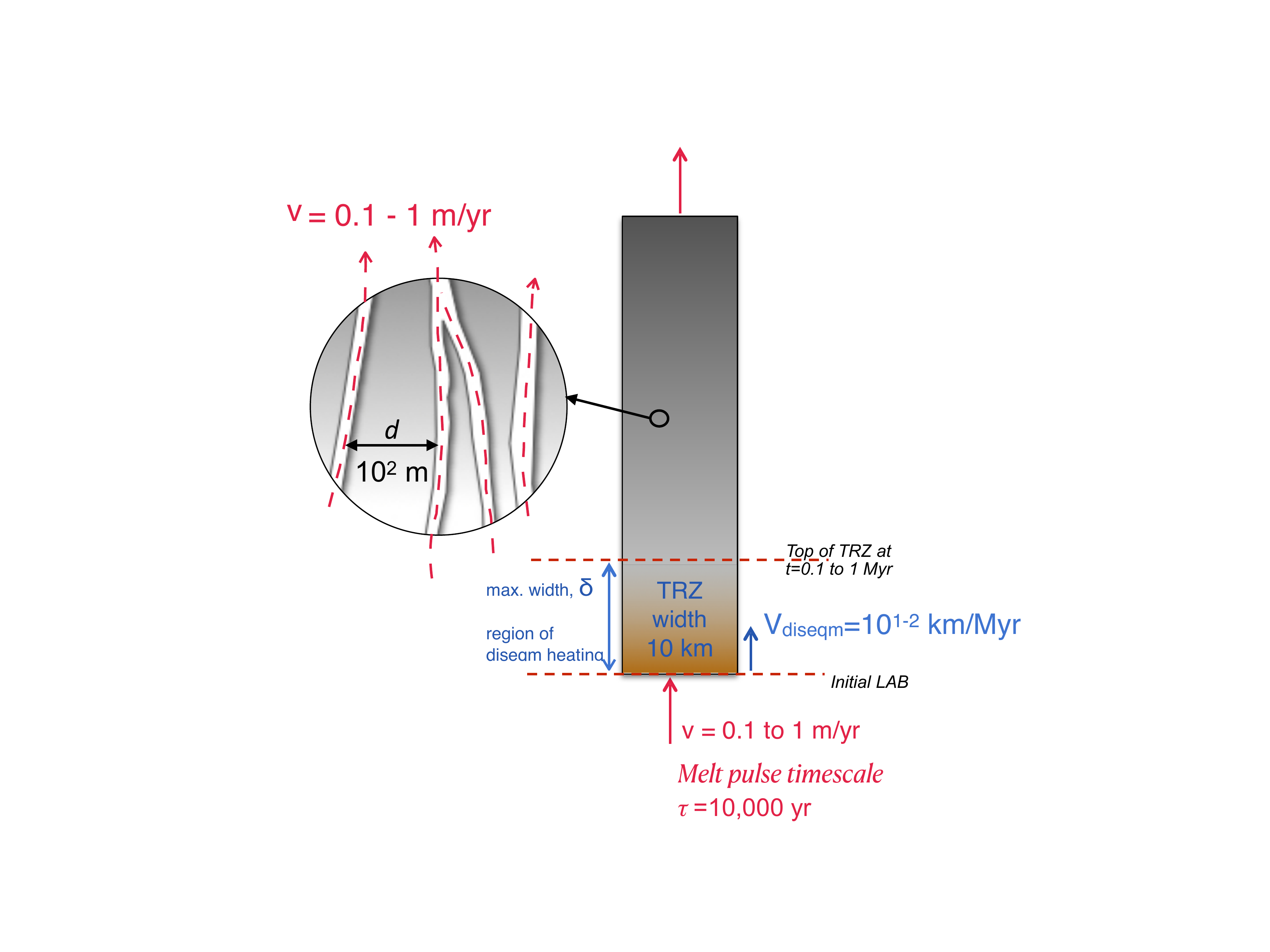}
\caption{Cartoon illustrating implications for a thermal re-working zone (TRZ) that forms a modified layer at the lowermost CLM as a result of disequilibrium heating. For episodic melt-infiltration into channels of spacing $d$$=$100 m, with period $\sim$10 Kyr, the TRZ is characterized by upward-decreasing degree of disequilibrium (indicated by the color). For in-channel material velocity $v$$= $0.1 to 1 m/yr, the TRZ grows to its steady-state width, $\delta$$ \sim$10 km, after 0.1 to 1 Myr.  Before reaching its final width, the TRZ may grow at a relatively large transient rate, $V_{diseqm}>$10 km/Myr.}
\label{fig:capstone}
\end{figure}

Given these limitations, the model above is a way to frame first-order questions and develop arguments related to the consequences of disequilibrium heating when it is dominated by downstream effects in the direction of transport.  
Taking the model domain to be analogous to the lowermost lithosphere, where melt or fluid transport may be channelized (Figure \ref{fig:capstone}), $x<0$ corresponds to a melt-rich sub-lithospheric region \cite<e.g. a decompaction layer,> {Holtzman:2010}, whereas the domain $x>0$ represents an initially sub-solidus lowermost CLM, and $x=0$ is the initial LAB (Figure \ref{fig:capstone}). 
Melt-infiltration into the lithosphere may be episodic, controlled by timescales associated with transport from the melt-generation zone to the LAB \cite<e.g.,>{stevensonmagmon84, spiegmagmon95}, processes of fracturing and crystallization in a diking boundary layer \cite<e.g.,>{Havlin2013} and melt supply from a deeper region of melt production \cite<e.g.,>{lambrift17}.
Although melt-infiltration at the base of the CLM is not expected to be periodic, the effects of a time-varying influx of hotter-than-ambient material within channels at the LAB may be understood in terms of the response to the equivalent sum of sinusoids.

Three key results emerge from the models above: (1) disequilibrium heating, estimated using the heat transfer coefficient, may be a significant portion of the heat budget at the LAB and the lowermost CLM, (2) a material-dependent velocity associated with transient disequilibrium heating, and (3) the existence of a thermal reworking zone (TRZ) associated with spatio-temporally varying disequilibrium heat exchange. 
Below I discuss each of these within the context of episodic melt-infiltration into the CLM in an intra-plate setting, specifically the Basin and Range province of the western US where deformation and 3D melt-rock interaction may be simplified by neglecting plate-boundary effects.
In this case, dominantly vertical heat transport within a slowly deforming lithosphere is a reasonable first-order assumption.

\textbf{i. Disequilibrium heating and the heat budget at the LAB.} 
The relative importance of disequilibrium heating at the LAB may be established by considering the effective heat transfer coefficient, $k$, and the factor which most strongly controls it, namely the average spacing of channels, $d$.  
For the material parameters in Table S1 (SI), and channel spacing of $d=1$ to 100 m, $k$ is in the range $k \approx 10^{-5}$ to $10^{3}$ W m$^{-3}$K$^{-1}$ (SI, Text S3).  
Physically, $k$ corresponds to fluid-solid heat transfer per unit time, per unit volume, per unit difference in temperature \cite{schumann1929heat}.
Therefore, for a 100 K excess temperature of the infiltrating melt, disequilibrium heating might contribute around $10^{-3}$ to $10^{5}$ W m$^{-3}$ to the heat budget at the LAB. 
This is a conservative estimate, given that the temperature difference between magma and the surrounding material may be larger (e.g., in Lherz the inferred contrast is $>$200 K \cite{soustelle2009}; and up to 1000 K in crust; \cite{lesherspera}).  
Similarly, plume excess temperatures are estimated to be as large as 250 K \cite{wangplume15}.

To put this in perspective, we now compare this estimated heat budget to the heat budget due to crystallization of melt in channels may be estimated using scaling arguments made in \citeA{Havlin2013}. 
Assuming that melt and rock are in equilibrium, \citeA{Havlin2013} estimate that the heat released by a crystallization front would contribute around $\rho H S_{dike}$, where $\rho$ is the melt density, $H$ is the latent heat of crystallization, $S_{dike}$ is a volumetric flow rate out of a decompacting melt-rich LAB boundary layer due to diking. 
For a representative porosity of $\phi=0.1$ within the dike, \citeA{Havlin2013} estimate $S_{dike}\approx 2\times 10^{-8}$ m$^3$/s.
Taking $\rho=3000$ kg m$^3$, and $H=3 \times 10^5$ J/kg, the heat source due to the moving crystallization front would be around $10^2$ W for each dike.  
If we assume that this heating takes place within a volume that is roughly the dike height $\times$ dike spacing $\times$ dike length, we can determine the power per unit volume generated due to crystallization.
For example, assuming dike heights of about $10^3$ m and dike spacing large enough for non-interacting dikes (as estimated by \citeA{Havlin2013}, a porosity of 0.1 would require a dike spacing of $\sim 10^3$ m), the heat source due to a crystallizing dike boundary layer would be $<10^{-4}$W/m$^3$ (per unit length along strike).  
These arguments corroborate the idea that disequilibrium heating during melt-rock interaction could be a significant portion of the heat budget at the LAB as compared to other expected processes, such as heating due to crystallization of melt in channels.

\textbf{ii. Progression of a disequilibrium heating zone/front at a rate $V_{diseqm}$.} 
The disequilibirum heating front is associated with a migration rate that is less than the in-channel material velocity, $v$. The importance of $V_{diseqm}<v$, is that $V_{diseqm}$ limits the rate at which the lowermost CLM may be modified by thermal disequilibrium during migration rate of either a disequilibrium front (Figure S2 a,b,c) or widening of a thermal reworking zone (Figure \ref{fig:sinecase}b).
It is important to note that $V_{diseqm}$ (Eqn \ref{eq:migrate}) is independent of temperature contrast between the CLM and infiltrating melt and depends only on the relative channel volume fraction, in-channel velocity and material properties. 
Assuming a channel volume fraction of 1 to 10 \% at the LAB, and material properties in Table S1 (SI), we would expect that $V_{diseqm}$ would be around 1 to 10 \% of the in-channel velocity (see SI, Text S2 and Figure S2, and Figure \ref{fig:capstone}).
For in-channel velocity in the range of 0.01 to 1 m/yr, we would predict that disequilibrium heating front at the LAB would migrate upward at a rate of $\approx$1 to 10$^2$ km/Myr, which is comparable to rates of CLM thinning predicted by heating due to the upward motion of a dike boundary layer (1 to 6 km/Myr in \citeA{Havlin2013}).
Interestingly, an upward-moving disequilibrium heating zone with $V_{diseqm}\approx 1$ to 10$^2$ km/Myr brackets the 10-20 km/Myr rate of upward migration of the LAB inferred from the pressure and temperature of last equilibration of Cenozoic basalts in the Big Pine volcanic Field in the western US \cite{plankforsyth16}.
An implication of the models here, therefore, is that disequilibrium heating may produce lithosphere modification at geologically-relevant spatial and temporal scales provided that the material velocity in channels at the LAB is on the order of $10^{-1}$ to 1 m/yr (Figure \ref{fig:capstone}).

\textbf{iii. Thermal reworking zone (TRZ).} A key result that may be relevant to the evolution of the LAB is that episodic infiltration of melts that are hotter than the surrounding CLM would lead to a long-lived region of disequilibrium heating within a thermal reworking zone or TRZ.
The TRZ would undergo a phase of transient widening (at a rate given by Eqn \ref{eq:migrate}), reaching a steady-state width $\delta$ that should scale as $\delta \sim \left[\phi v_{chan}d^{-2}\tau^{2} \right]$ where $d$ is a characteristic scale of channelization and $\tau$ is a timescale associated with the episodicity of melt-infiltration (Figures \ref{fig:sinecase}d and \ref{fig:capstone}).
This scaling gives us a way to conceptualize the modification of the lowermost CLM as a zone that may encompass a variable thickness TRZ, depending on variability in $v$ and in the timescale of melt-infiltration (Figure \ref{fig:capstone}).
Regions where the timescale of episodic melt-infiltration is longer are predicted to have a thicker zone of modification at the LAB.
For example, for a channel spacing of $d = 10^{2}$ m, disequilibrium heating by repeated melt pulses that last around 10 Kyrs implies a maximum thickness of roughly 10 km for the zone of modification (Figure \ref{fig:sinecase}d).
In this scenario, the TRZ grows to this maximum width over a timescale governed by $\delta/V_{diseqm}$; for $V_{diseqm}=10$ km/Myr, which corresponds to melt velocity of roughly 0.1 m/yr (see (ii) above), the 10 km wide TRZ would be established within about 1 Myr (Figure \ref{fig:capstone}), comparable to rates of CLM modification inferred from observations in \citeA{plankforsyth16}.

These scaling arguments lead to the idea that perhaps the TRZ represents a zone of thermal modification at the base of the CLM that may also correspond to (or encompass) a zone of rheologic weakening and/or in-situ melting if the infiltrating fluids are hotter than the ambient material. 
The dynamic evolution of the LAB during episodic pulses of melt-infiltration is beyond the scope of the simple models above (which assume a stationary, undeforming matrix).
However, assuming mantle material that obeys a temperature and pressure-dependent viscosity scaling relation such as in \citeA{Hirth:2003p132}, at an LAB depth of about 75 km where we assume that the ambient mantle is cooler than the dry solidus (e.g., 1100$^o$C + 3.5$^o$C/km; \citeA{plankforsyth16}), we would expect a significant viscosity reduction during heating (e.g., factor $\approx1/62$ for a temperature increase of 100 K).
This effect is weaker, but still important for a deeper LAB; e.g, at 125 km depth, the viscosity reduction would be a factor $\approx1/18$ for a temperature increase of 100 K.

Interestingly, geochemical evidence from Cenozoic basalts from the western US, particularly space-time variations in volcanic rock Ta/Th and Nd isotopic compositions suggest that the timescale of modification and removal of the lowermost CLM is on the order of 10$^1$ Myrs \cite{farmerTaTh20}. 
These authors argue that the observed transition from low to intermediate to high Ta/Th ratios indicates a change from: arc/subduction-related magmatism, to magmatism associated with in-situ melting of a metasomatized CLM (the ``ignimbrite flare-up"), to magmatism due to decompression and upwelling after removal of the lowermost CLM.
At a minimum, the observed timescale of the transition in Ta/Th ratios in volcanic rocks (10$^1$ Myrs) in the western US should be comparable to the timescales of degradation of the CLM.
If correct, these interpretations and observations are promising and provide an important avenue for exploring the role of thermal and chemical disequilibrium during melt-rock interaction and destabilization of the CLM in an intra-plate setting.

\section{Conclusions}

In summary, I have presented arguments supporting the role of disequilibrium heating in the modification of the base of the CLM during melt-infiltration into and across the LAB.
Infiltration of pulses of hotter-than-ambient material into the LAB should establish a thermal reworking zone (TRZ) associated with disequilbrium heat exchange. 
The spatial and temporal scales associated with the establishment of the TRZ are comparable to those for CLM modification inferred from geochemical and petrologic observations intra-plate settings, e.g., the western US.
Disequilibrium heating may contribute more than $10^{-3}$W/m$^3$ to the heat-budget at the LAB and, for transport velocity of 0.1 to 1 m/yr in channels that are roughly 10$^2$ m apart, a 10 km wide TRZ may be established within 1 Myr.
Disequilibrium heating during melt-infiltration may be an important process for modifying the lowermost CLM and may play a role in the rheologic weakening preceding mobilization (and possibly removal) of the lowermost CLM. 

\newpage
\section{Supplementary Information}
\noindent\textbf{Text S1. Geologic evidence for the role of thermal disequilibrium in the lower continental lithosphere from the Lherz and Ronda peridotite massifs}

Two important conclusions relevant to this work emerge from studies in the Lherz and Ronda peridotite massifs:  (1) First, we now know that ``lherzolite" (named after its type-section in the Lherz massif), commonly regarded as pristine, fertile sub-continental lithospheric mantle, is actually derived from refertilization of a depleted, harzburigitic parent \cite<e.g.,>[]{leroux2007lherz, leroux08isotope}; (2) Second, careful microstructural, geochemical and petrologic work has documented the dominant effect of a steep thermal gradient associated with the region of contact and interaction between partial-melt-rich (sub-lithospheric) regions and the lithosphere.  

The importance of thermal disequilibrium in the lower lithosphere is most clearly demonstrated by in Ronda \cite{soustelle2009}. These workers provide a quantitative estimate of this transient thermal gradient ($\approx 230 ^{o}$C/km, or more than an order of magnitude larger than a typical equilibrium geothermal gradient expected at the LAB). 
\citeA{soustelle2009} show that thermal disequilibrium heating also drove partial-melting of the lithosphere above/around the melt-rich region.  
Indeed, the authors recognize this as a transient LAB and coin the term "asthenospherization" for the thermally-controlled disequilibrium processes, including heating. 
The spatial scale over which this disequilibrium heating is observed in Ronda ($\sim 1$ km) forms a constraint used here.

Interpreting the observed region of thermal disequilibrium as part of the TRZ, I explore what timescales of melt-infiltration give rise to TRZ widths on the order of 1 km (Figure 2c).
The horizontal lines in Figure 2c denote the timescales of melt-infiltration that would be required to give rise to a $10^{0}$ to $10^{1}$ km region where thermal disequilibrium may be important, as observed. 
%

\newpage
\noindent\textbf{Text S2. Nondimensional system}

To non-dimensionalize the system of equations 1 and 2 \cite<e.g.,>{spiga81}, we define the normalized relative temperature, $T'_f=(T_f-T_0)/\Delta T$ and $T'_s=(T_s-T_0)/ \Delta T$, where $T_0$ is reference temperature and $\Delta T$ is a temperature perturbation (described below).
We also introduce the dimensionless position, $x'=xk_f/v$, a dimensionless time, $t'=k_st$, and the heat capacitance ratio $z=k_s/k_f=\phi c_f/(1-\phi)c_s$. 
The non-dimensional versions of equations 1 and 2 are now (\ref{eq:Tfnd}) and (\ref{eq:Tsnd}):
\begin{linenomath*}
	\begin{equation}
		z\frac{\partial{{T'}_f}}{\partial{t'}}+ \frac{\partial{{T'}_f}}{\partial{x'}}=-({T'}_f - {T'}_s)
		\label{eq:Tfnd}
		\tag{S1}
	\end{equation}
	\begin{equation}
		\frac{\partial{{T'}_s}}{\partial{t'}}=({T'}_f - {T'}_s)
		\label{eq:Tsnd}
		\tag{S2}
	\end{equation}
\end{linenomath*}
(see also \citeA{spiga81}).  
It is clear that for a given temperature difference, $({T'}_f - {T'}_s)$, the behavior of this system is governed by $z$ ($1/z$ is a dimensionless velocity).

Analytic solutions for this set of equations have been derived for a number of limiting cases, particularly for large $k$ \cite{spiga81, kuznetsov94, kuznetsov95,kuznetsov95b,kuznetsov96}, and were used to benchmark the numerical calculations in this study. 


\newpage
\noindent\textbf{Text S3. Heat transfer coefficient, $k$, and relation to channel spacing, $d$}

The factors that determine $k$ can be illustrated by considering that heat transfer rate across channel walls must depend on the  geometry of walls and also on the effective thermal conductivity of the channelized domain.  
Although the geometry of the channels may be complex, this model considers one aspect of it: the specific wall surface area (wall area per unit volume), $a_{sf}$, which is a function of the length-scale of channelization.  
In the grain-scale porous flow case considered in \citeA{schumann1929heat} for example, if the solid matrix is made of spheres with an average particle diameter $d$, then the specific area for a grain is $S_0 = 6/d$, so $a_{sf}=S_0(1-\phi) = 6(1-\phi)/d$ \cite{dullien79}.  
This sets a limit for channels, where we shall assume that the specific surface area is $a_{sf}\sim A(1-\phi)/d$, where $A$ is a number that is between 2 (as for a single cylindrical channel with small volume fraction $\phi$) and 6 \cite{dixoncresswell79}.  
Whereas the specific wall area is a geometric factor, the effective conductivity of the medium depends on the Nusselt number, $Nu$.
Theoretical arguments in \citeA{dixoncresswell79} show that the effective thermal conductivity may be written in terms of the individual in-channel and out-of-channel thermal conductivities $\lambda_{f}$ and $\lambda_{s}$
(basically taking the channels and non-channel regions in parallel):
\begin{linenomath*}
	\begin{equation}
		\frac{1}{C_{eff}}=\left[\frac{1}{{Nu}\lambda_f}- \frac{1}{\beta\lambda_s}\right]
		\label{eq:keff}\tag{S3}
	\end{equation}
\end{linenomath*}
where $\beta=10$ for spherical matrix grains, 8 for cylinders, and 6 for slabs \cite{dixoncresswell79}.  
Therefore, the range of $\beta=$6 to 10 represents the highly-channelized vs porous flow end-member geometries.  
For slow flows (Reynolds number Re $\ll100$), \citeA{handleyheggs68} argue that ${Nu}$ ranges from $0.1$ to $12.4$ \cite{dixoncresswell79} (Table 1 in text).  
The relevant quantity that determines $k$ is an effective ``conductance'' $C_{eff}/d$, so that $k=C_{eff}a_{sf}/d$,
\begin{linenomath*}
	\begin{equation}
		k=\frac{1}{d}\left[\frac{1}{{Nu}\lambda_f}- \frac{1}{\beta\lambda_s}\right]^{-1}\frac{A(1-\phi)}{d}
		\label{eq:kschum}\tag{S4}
	\end{equation}
\end{linenomath*}
a product of a material-dependent quantity and a geometry-dependent quantity.  

\begin{figure}[!htb]
\centering
\begin{tabular}{c  c}
	\includegraphics[trim =1mm 1mm 1mm 1mm, clip, width=0.38\textwidth]{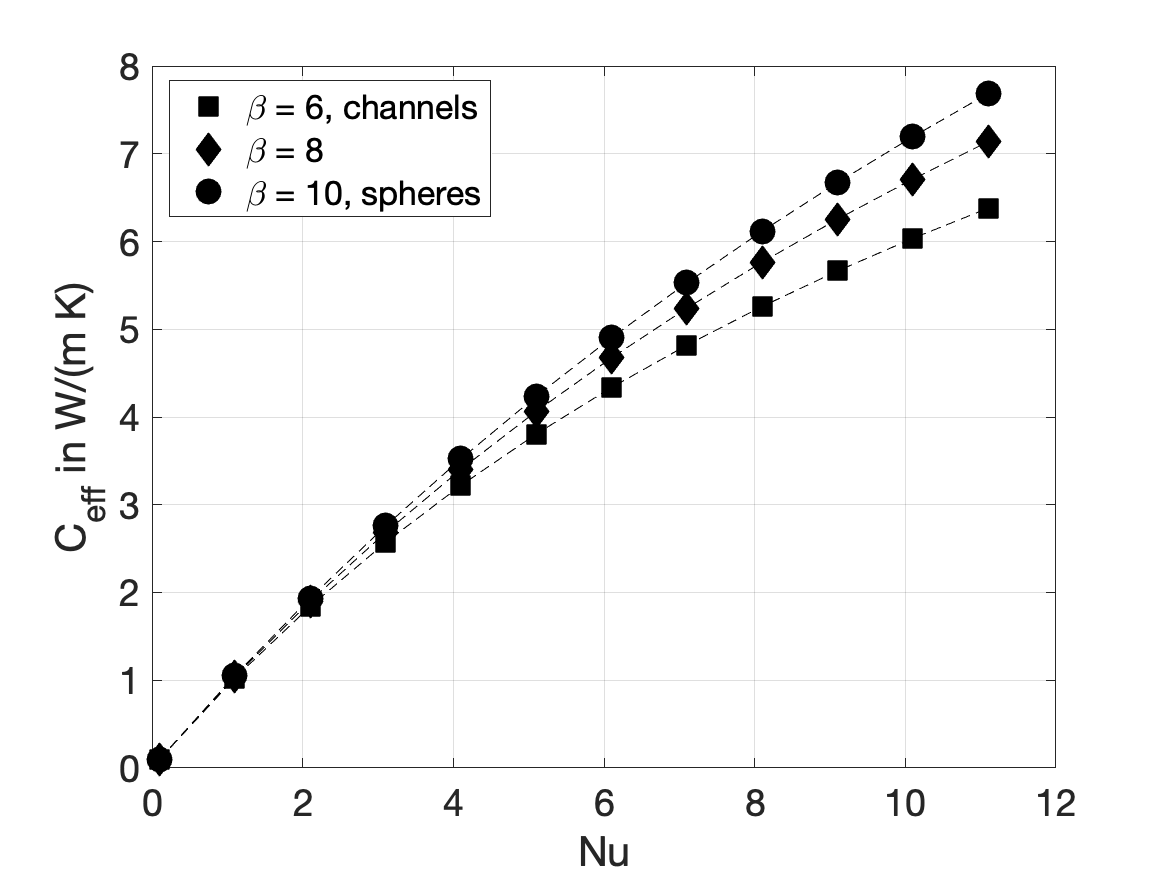} &
	\includegraphics[trim =1mm 1mm 1mm 1mm, clip, width=0.38\textwidth]{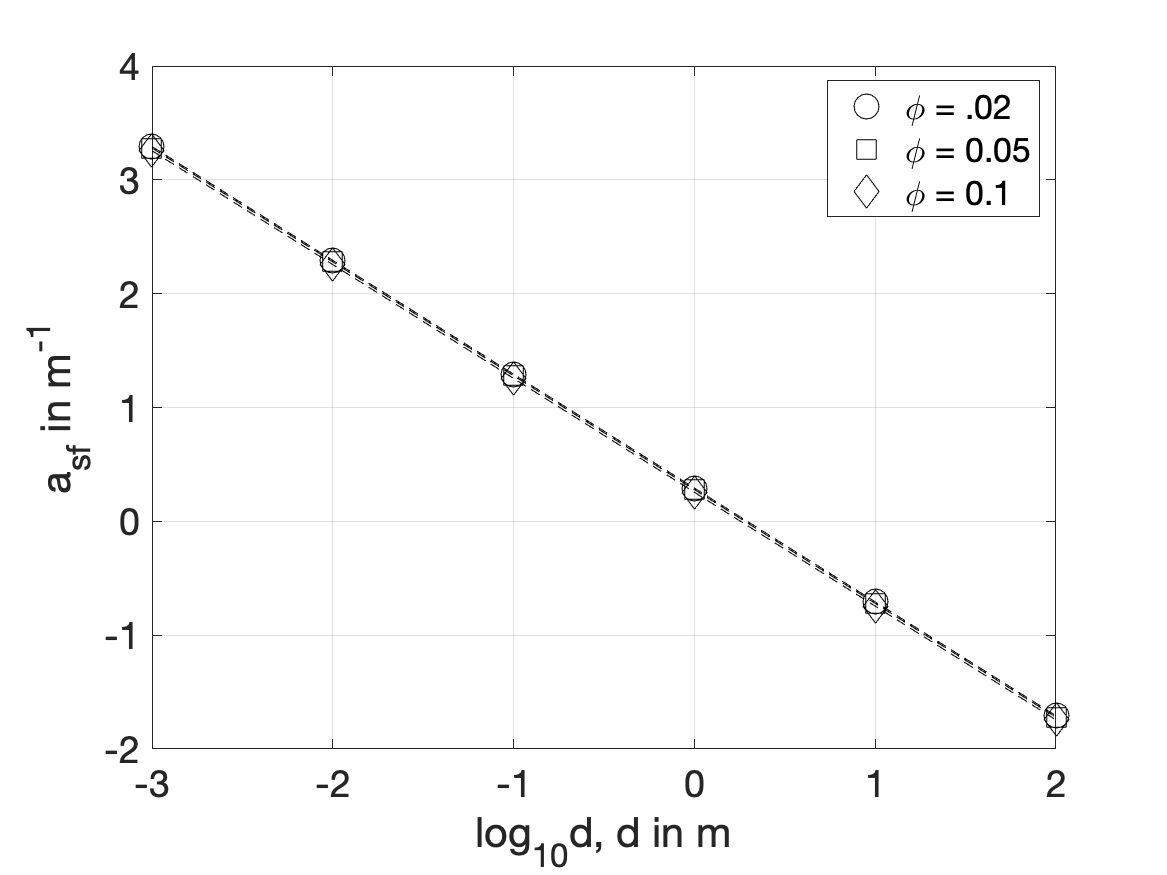} \\ 
	(a) & (b) \\
\end{tabular}
\begin{center}
	\includegraphics[trim =1mm 1mm 1mm 1mm, clip, width=0.38\textwidth]{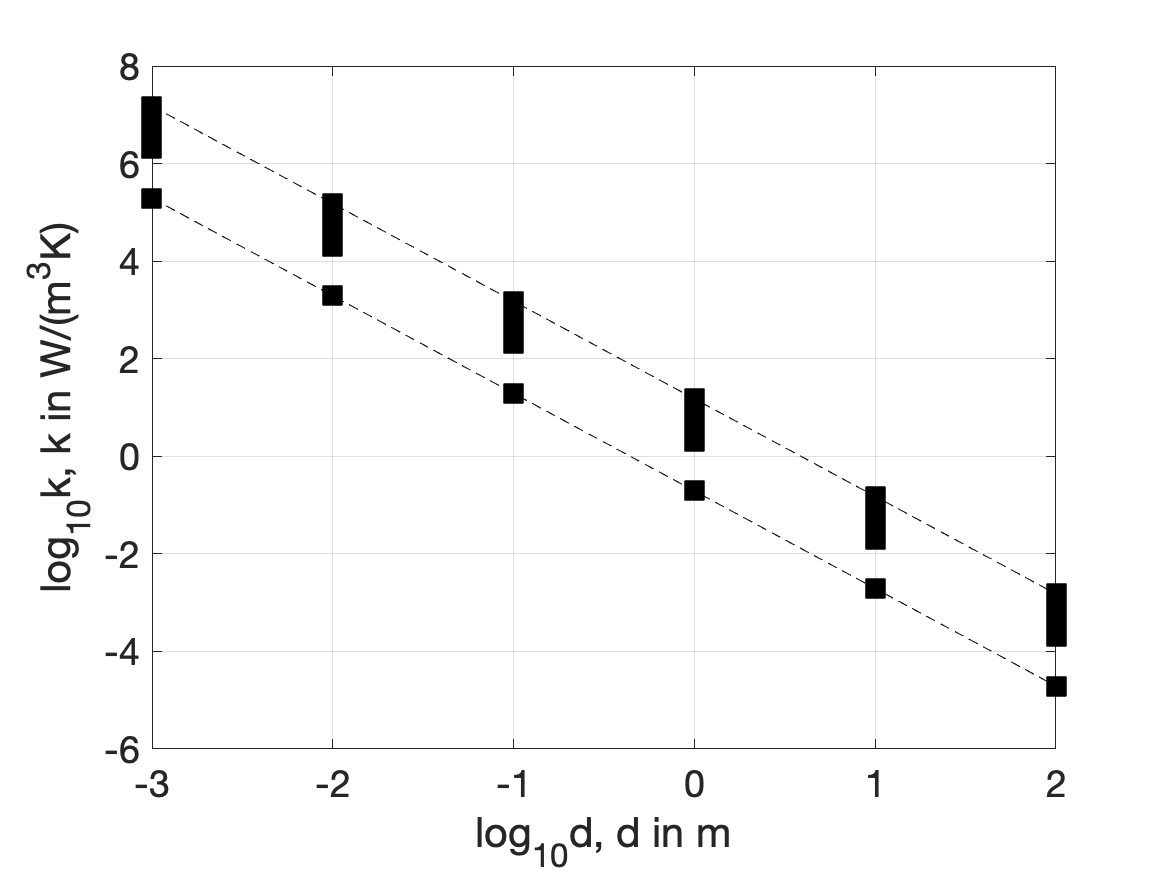} \\
	(c)\\
\end{center}
\caption{(a) Effective thermal conductivity, $C_{eff}$ in Eqn \ref{eq:keff}, as a function of Nusselt number, (b) geometric factor, $a_{sf}$, as a function of channelization scale, $d$, and (c) heat transfer coefficient, $k$, as a function of channelization scale $d$.  For a fixed $d$, the dashed lines in (c) delineate the variation in $k$ for the range of $\beta$ values in (a) and $\phi$ values in (b), illustrating that $k$ is mainly controlled by $d$, rather than the other parameters.}
\label{fig:kfig}
\end{figure}

Turning now to physical properties relevant to the transport of melts through the lithosphere, the channelized domain may be thought of as consisting of a mixture of melt+grains throughout, but with variable grain-scale porosity $\varphi$.
Specifically, the fluid-rich channels would have a higher $\varphi_{f}$ than the surroundings $\varphi_{s}$. 
The thermal conductivity inside and outside the channels would then be a volume average in each, e.g., inside channels, $\lambda_f = \varphi_{f}\lambda_{melt} + (1-\varphi_{f})\lambda_{grain}$, where $\lambda_{melt}$ and $\lambda_{grain}$ are the values for, say basaltic melt and peridotitic grains.
Similarly, outside channels the volume average would be $\lambda_s = \varphi_{s}\lambda_{melt} + (1-\varphi_{s})\lambda_{grain}$.
Here I do not specify reasonable ranges of $\varphi_{in}$ and $\varphi_{out}$, but rather I focus on determining an upper limit to the role of thermal disequilibrium across channel walls.
Therefore, to explore the end-member case, I take $\varphi_{f}=1$ and $\varphi_{s}=0$, so that $\lambda_f=\lambda_{melt}$ and $\lambda_s=\lambda_{grain}$.
Using a reasonable conductivity for basaltic magma of $\lambda_{melt}=1$ W/(m K) \cite{lesherspera}, $\lambda_{grain}=2.5$ W/(m K) for the solid grains, and taking $Nu=0.1$ to $12.4$, $\beta=6$, 8, or 10, and $A=6$, we find that the effective conductivity $C_{eff}$, $a_{sf}$ and $k$ are within the ranges shown in Figure \ref{fig:kfig}.
(Note that choosing $A\approx2$, for a more channelized geometry, would change $k$ by less than an order of magnitude.) 
Specifically, Eqn (\ref{eq:kschum}) shows that $k\sim d^{-2}$ and strongly decreases with increasing spatial scale of channels; Figure \ref{fig:kfig}c.

\newpage
\noindent\textbf{Text S4. Response to a step-function}
As material with a perturbed temperature enters channels at $x=0$, the surroundings heat up while the fluid-rich channels cool (for positive perturbation $\Delta T$). 
The perturbation ``front'', the farthest extent of channel material with perturbed temperature, is at $x_{\text{front}}=vt$ (or $x'_{\text{front}}=t'/z$; where the dimensionless velocity inside channels is $1/z$; see Text S2).  

Non-dimensional equations (\ref{eq:Tfnd}) and (\ref{eq:Tsnd})in Text S2 may be solved for the thermal evolution subject to the initial and boundary condition: ${T'}_s={T'_f}=0$ initially and ${T'}_f(x=0, t\geq0)=1$.  
The behavior of the nondimensional system is controlled by $z$, the dimensionless in-channel velocity, which is a function of both material properties $c_s$ and $c_f$ and the channel volume fraction, $\phi$.
To be consistent with the end-member case discussed in Text S2, where in-channel grain-scale porosity is taken to be 1 and the out-of-channel is 0, I use heat capacitances for basaltic melt and peridotitic  grains (Table S1).
Since the material properties $c_s$ and $c_f$ are held constant, there is a unique mapping between $z$ and $\phi$, the channel volume fraction (we use $\phi=$ 0.01 to $0.20$, corresponding to values of $z=0.0096$ to $0.2376$; Table S1 (SI).

\begin{figure}[!htb] 
\centering\includegraphics[trim =55mm 15mm 40mm 16mm, clip, width=0.6\textwidth]{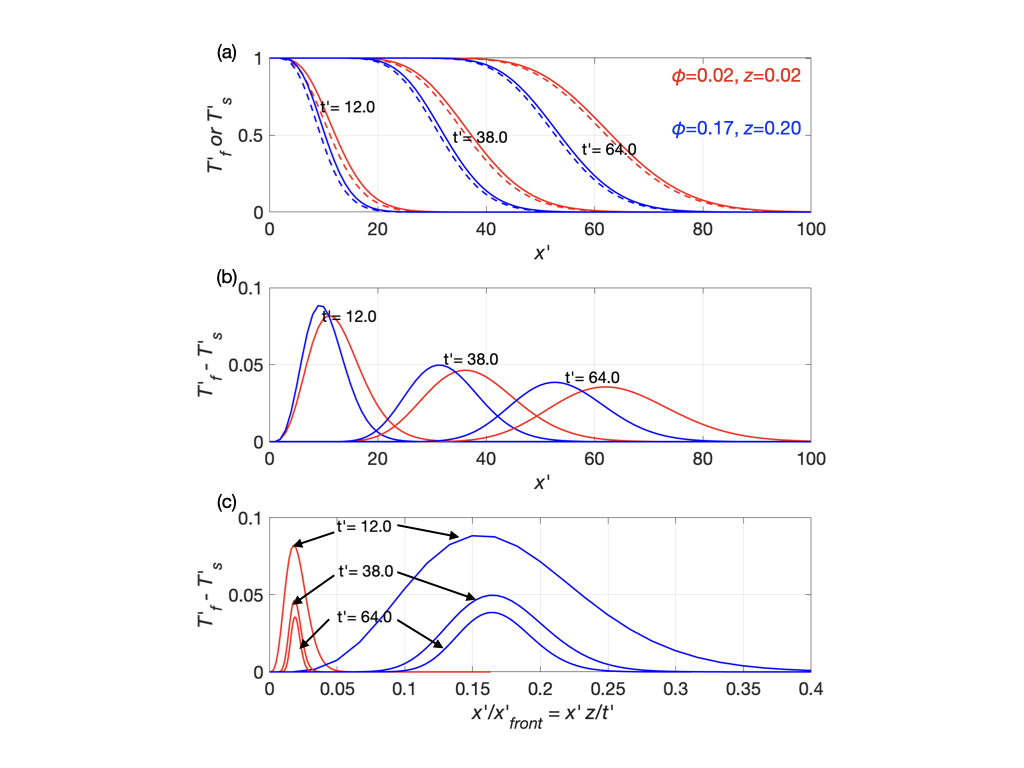} 
\includegraphics[trim =8mm 2mm 12mm 8mm, clip, width=0.38\textwidth]{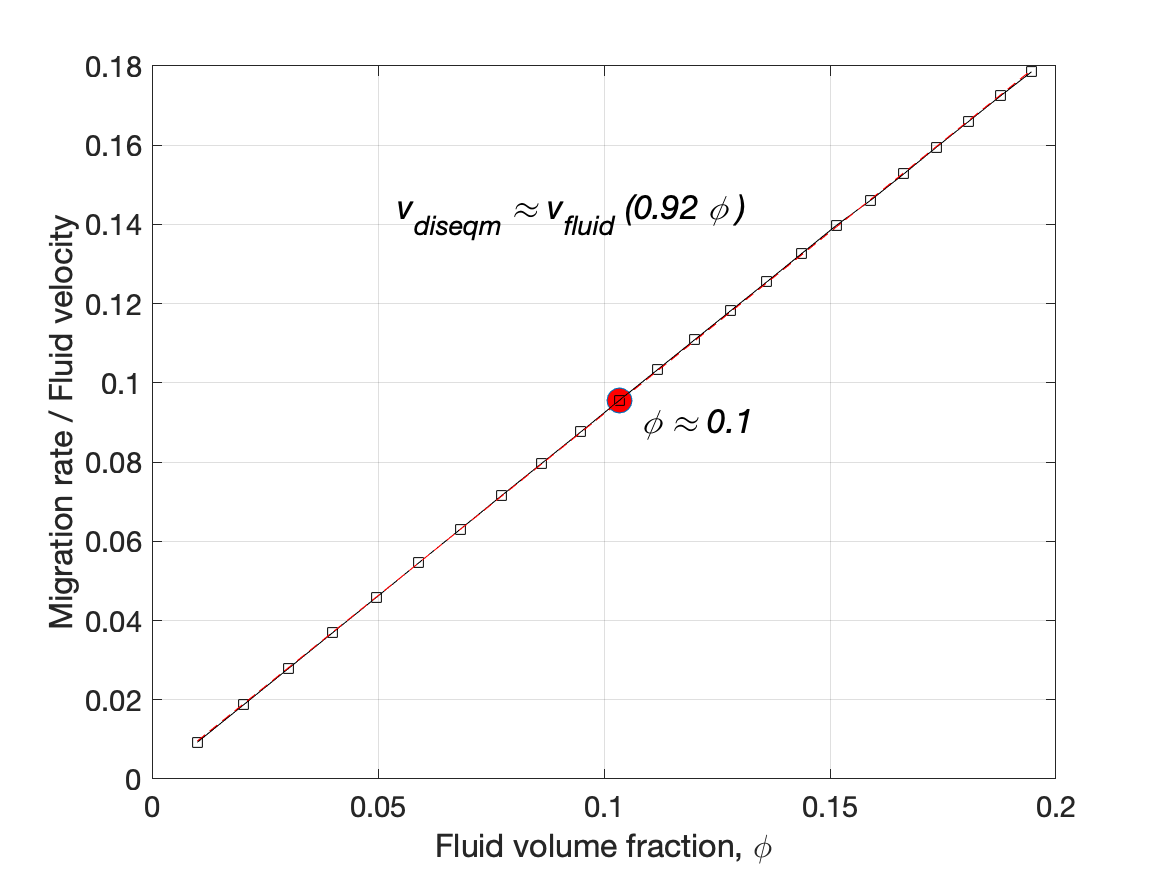} 
\caption{(a) Normalized temperature as a within channels (solid lines) and surroundings (dashed lines) at different values of the dimensionless time, $t'$, since a step function perturbation, as indicated.  The colors represent two different channel volume fractions, $\phi$, and therefore different $z$. (b) Normalized temperature difference across channel walls as a function of dimensionless position, shown for the cases considered in (a).  (c) The same profiles as in (b), but now plotted as a function of position normalized by the perturbation front location, $x'_{\text{front}}$=$t'/z$; stationarity of the disequilibrium zone in this plot indicates that the disequilibrium zone migrates at a constant, z-dependent fraction of the in-channel velocity. (d) Normalized migration rate of the zone of disequilibrium as a function of channel  volume fraction, $\phi$. Red dot is for $\phi\approx$ 0.1, corresponding to models shown in Figure \ref{fig:front} and Figure 2 in text.}
\label{fig:diffz}
\end{figure}

As we might expect, the behavior of the nondimensional system (Eqns 3 and 4 in text) is governed by $z$, the heat capacitance ratio.
The response to a step-function is essentially a transient disequilibirum front, traveling inward at $V_{diseqm}$, behind which the solid and fluid equilibrate at the new inlet temperature.  
In response to a step-function increase in the inlet temperature, temperature profiles within the domain exhibit a transition or disequilibrium zone, lagging behind the fluid front (Figure \ref{fig:diffz}). 
Ahead of this disequilibrium zone, the channels are in equilibrium with the surroundings at the initial ambient temperature, ${T'}_s={T'_f}=0$.
Behind this zone, the channels are in equilibrium with the surroundings at the inlet temperature, ${T'}_s={T'_f}=1$.
%

%

%
Following an initial lag time (when the maximum disequilibrium is at $x'=0$), the disequilibrium zone migrates inward migration at a steady speed, a fixed fraction of the in-channel velocity, $v$ (Figure \ref{fig:diffz}c and \ref{fig:diffz}d).
The ratio of the migration rate of the disequilibrium zone to $v$ is controlled by the heat capacity ratio, $z$, and therefore by the channel volume fraction, $\phi$ (Figure \ref{fig:diffz}).
In the near-equilibrium limit, ($T'_f-{T'_s}\approx0$), \citeA{kuznetsov94} shows that the shape of the temperature difference function (Figure \ref{fig:diffz}b) approaches a Gaussian with width that depends on $\sqrt{t'}$ and the zone of disequilibrium migrates at speed $v {c_{f}}/({\phi c_{f}+ (1-\phi)c_{s}})$.
Our models show that, when there is significant disequilibrium, the zone of disequilibrium migrates with a rate given by Eqn 3 in text (Figure \ref{fig:diffz}d), which does not depend on $k$, the heat transfer coefficient.

\begin{figure}[!htb] 
\centering
\includegraphics[trim =15mm 38mm 16mm 18mm, clip, width=0.9\textwidth]{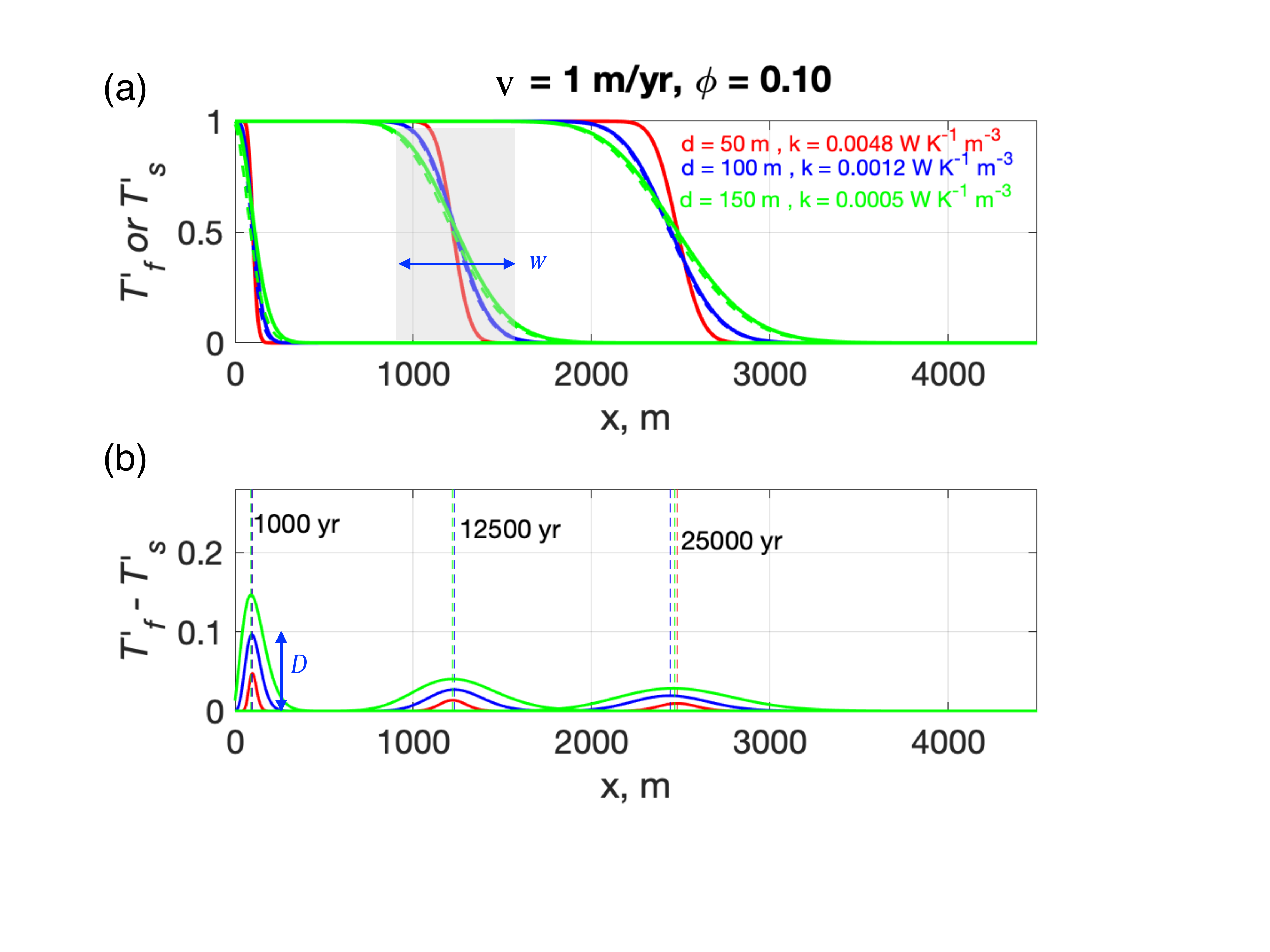}  
\begin{tabular}{cc}
	\includegraphics[trim =1mm 1mm 1mm 1mm, clip, width=0.38\textwidth]{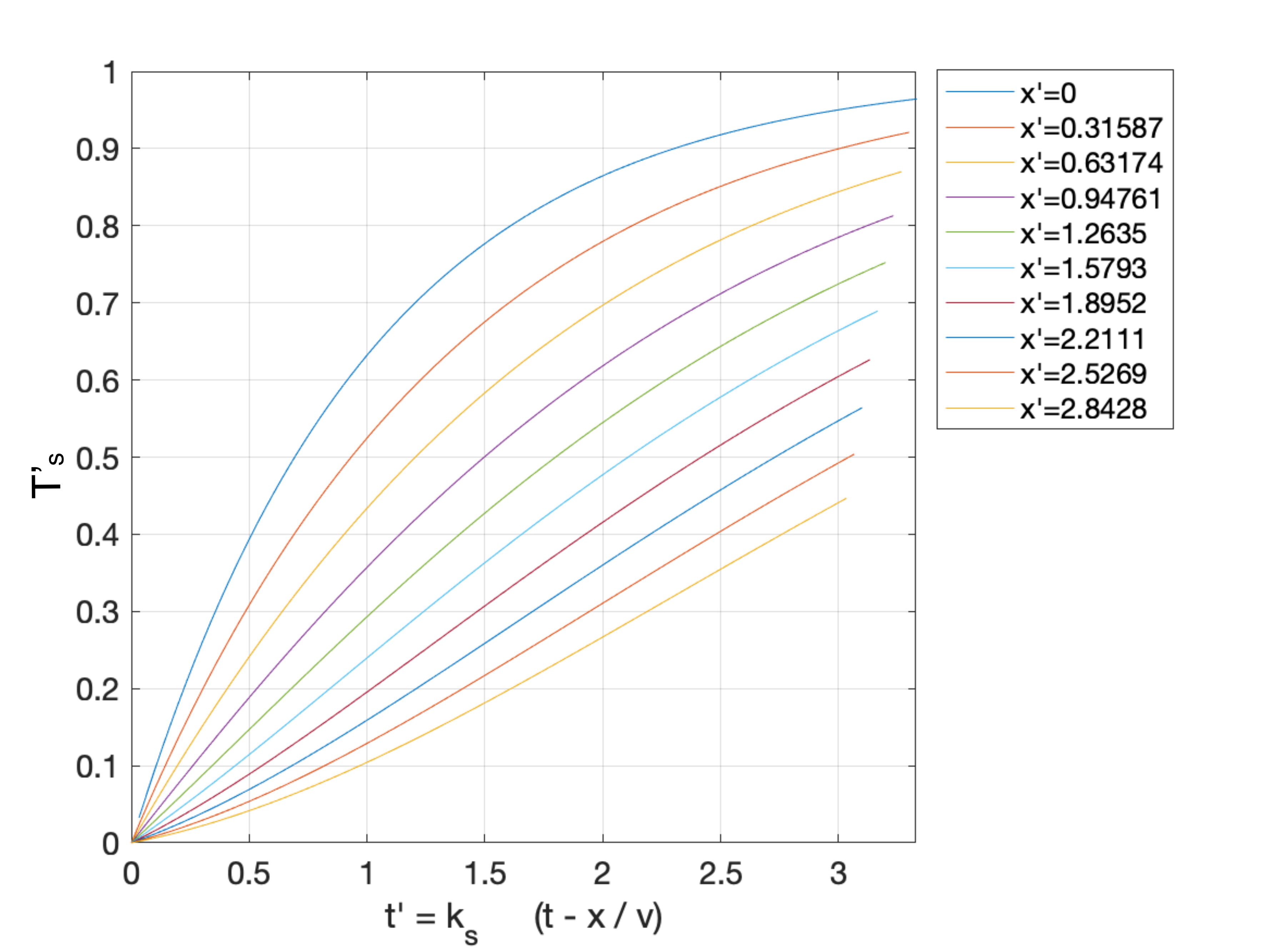} &
	\includegraphics[trim =1mm 1mm 1mm 1mm, clip, width=0.38\textwidth]{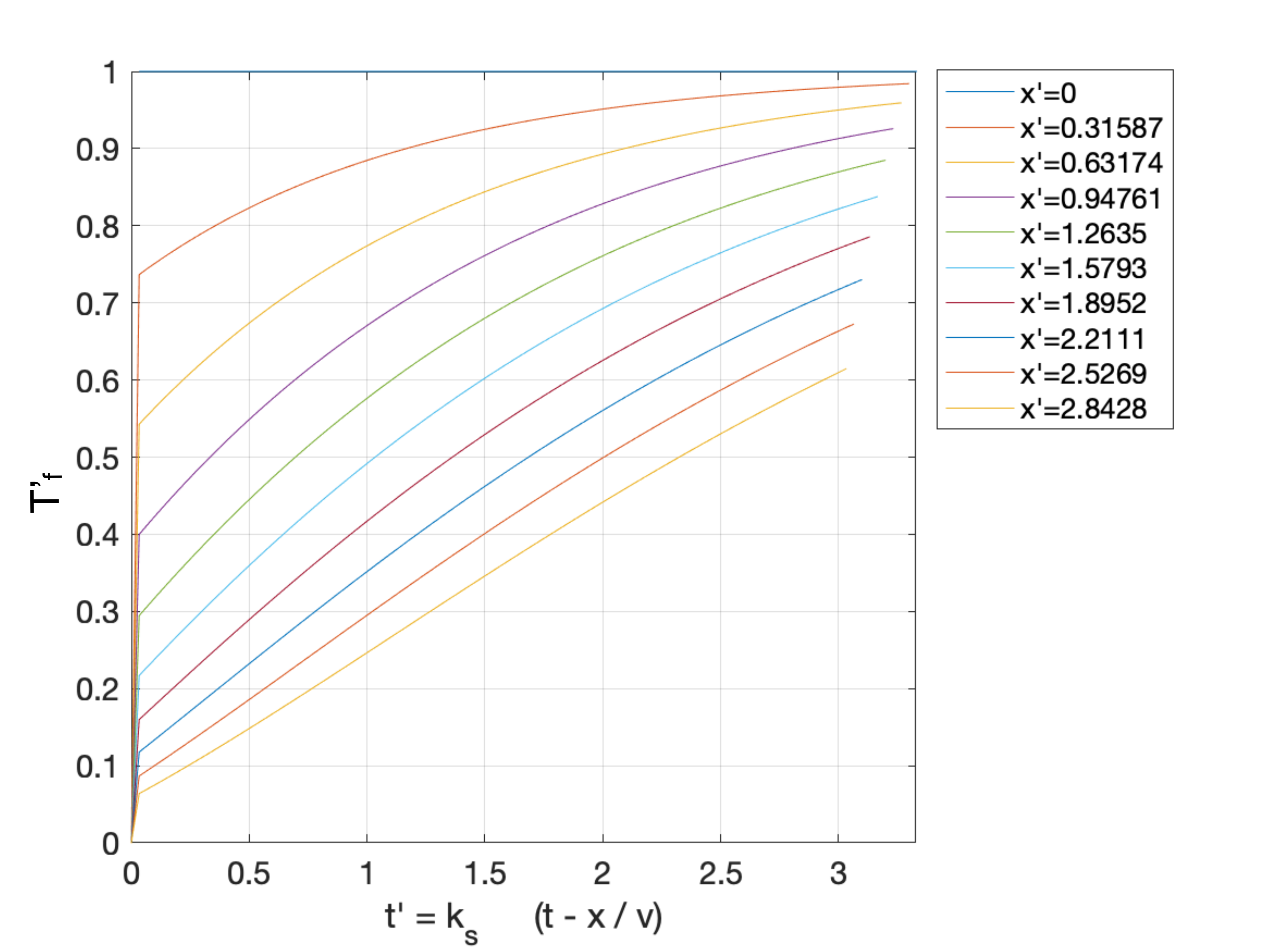} \\
\end{tabular}
\caption{\linespread{1}{{(a) Normalized temperature profiles, $T'_{s}$ (dashed) and $T'_{f}$ (solid) for in-channel velocity $v=1$ m/yr, at times $t=1$, $12.5$, and $25$ Kyr.  The temperature profiles transition between the incoming channel material temperature (=$1$, left) and the initial ambient temperature (=$0$, right), for two cases with different channel spacing, $d$, and heat transfer coefficient, $k$, as indicated.  The transition region (e.g., highlighted in gray at $t$=$12.5$ Kyr), has width, $w$, that is larger for smaller $k$ (large $d$) and increases over time.  (b) The degree of disequilibrium is characterized by the maximum difference across channel walls, $D$.  $D$ is greater for smaller $k$ and decreases as a function of time. (c) $T'_{s}$ and (d) $T'_{f}$ as a function of dimensionless time since first contact with the perturbation front.}}}
\label{fig:front}
\end{figure}

Although $V_{diseqm}$ is independent of $k$, the degree of disequilibirum is not.  
We illustrate the dependence on $k$ for the specific case where the in-channel velocity $v$=1 m/yr, and consider channel spacings, $d=50$ to $150$ m, which correspond to $k=5\times 10^{-3}$ and $5\times 10^{-4}$ W m$^{-3} $ K$^{-1}$, respectively (Figure \ref{fig:front}). 

The migration of the disequilibrium front may be thought of as the motion of the locus of maximum heating, which moves at speed $V_{diseqm}\approx v/10$, for $\phi=0.1$ in Figure \ref{fig:front}.
The degree of disequilibrium within the migrating disequilibrium zone, $D$, decays as $1/\sqrt{t}$ \cite<e.g.,>{kuznetsov94} and is controlled by the heat transfer coefficient, $k$.  $D$ scales as $1/\sqrt{k}$ and therefore is a linear function of $d$, the channel spacing (Figure \ref{fig:front}b).
This transient is apparent in temperature-time paths (Figure \ref{fig:front}c and \ref{fig:front}d), where the approach to steady-state occurs on a timescale governed by $\phi v/k$.

\newpage
%
%

\begin{table}[h!]
	\renewcommand{\arraystretch}{1.1}
	\caption{Table S1. Material properties and constants used in calculations}
	\label{tab:props}
	\small
	\centering
	\begin{tabular}{p{5.5cm}p{2 cm}p{3cm}p{5.5cm}} 
		\textbf{Name} & \textbf{Symbol} & \textbf{Value or range}  & \textbf{Source/Comments}\\
		\hline
		\rowcolor{Gray}
		Melt, grain density &  $\rho_{melt}$, $\rho_{grain}$ & 2800, 3300 kg/m$^3$ & \citeA{lesherspera}\\
		Melt specific heat capacity  &  $c_{p\it {melt}}$ & 1400 J/(kg K) & \citeA{lesherspera}\\
		\rowcolor{Gray}      Grain specific heat capacity  &  $c_{p\it {grain}}$ & 1250 J/(kg K) & \citeA{lesherspera}\\
		Melt heat capacity per volume & $c_{melt}$ & $3.920\times10^6$ J/(m$^3$ K) & $c_{p\it{melt}}\times\rho_{melt}$\\
		\rowcolor{Gray}
		Grain heat capacity per volume & $c_{grain}$ & $4.125\times10^6$ J/(m$^3$ K) & $c_{p\it{grain}}\times\rho_{melt}$\\
		In-channel, out-of-channel grain-scale porosity &  $\varphi_{f}$, $\varphi_{s}$ & 1, 0 & End-member case maximizing material property contrast (Text S3)\\
		\rowcolor{Gray}
		In-channel heat capacity per volume & $c_f$ &  $3.920\times10^6$ J/(m$^3$ K) & $c_f= \varphi_f c_{melt}+(1-\varphi_f)c_{grain}=c_{melt}$\\
		
		Out-of-channel heat capacity per volume & $c_s$ & $4.125\times10^6$ J/(m$^3$ K) & $c_s= \varphi_s c_{melt}+(1-\varphi_s)c_{grain}=c_{grain}$\\
		\rowcolor{Gray}
		Heat transfer coefficient & $k$  &$10^{-5}$ to $10^{1}$ W/m$^{3}$K & this work (SI, Text S1)\\
		
		Channel volume fraction  &  $\phi$ & 0.1 to 0.2 &  \citeA<e.g.,>{pec2017} \\
		\rowcolor{Gray}
		Channel average (linear) velocity relative to surroundings  &  $v$ & 1 to 100 mm/yr & \citeA<e.g.,>{rutherford08} \\
		
		Weighted heat capacity ratio  & $z$ & 0.0096 to 0.2376 & calculated\\
		\rowcolor{Gray}
		Fluid-solid Nusselt number & $Nu$ & $0.1$ to $12.4$ & for slow flows \citeA{handleyheggs68} \\
		
		Constant in Eqn \ref{eq:keff} & $\beta$ & 6 to 10 & \citeA{dixoncresswell79}\\
		\rowcolor{Gray}
		Constant in Eqn \ref{eq:kschum} & $A$ & 2 to 6 & \citeA{dullien79}\\
		
		Separation of fluid-rich channels & $d$ & $10^{-1}$ to $10^2$ m & \citeA{leroux08isotope}\\
		\hline
	\end{tabular}
\end{table}

\newpage


%
%
%
%
%
\newpage

\newpage

\newpage

\bibliography{/Users/mousumi/Desktop/MyRefs_global}

\end{document}